\documentclass[aps,prc,twocolumn,superscriptaddress,showpacs,nofootinbib]{revtex4-2}

\usepackage{amsmath,amssymb}

\usepackage{bm}
\usepackage{graphicx}
\usepackage{color}

\usepackage{epsfig}
\usepackage{graphicx}
\usepackage{dcolumn}
\usepackage{bm}
\usepackage{multirow}
\usepackage{array}
\usepackage{slashed}
\usepackage{amsmath}
\usepackage[titletoc]{appendix}
\usepackage{color}
\usepackage{diagbox}
\usepackage[colorlinks,citecolor=blue]{hyperref}
\newcommand{\orcid}[1]{\href{https://orcid.org/#1}{\includegraphics[width=9pt]{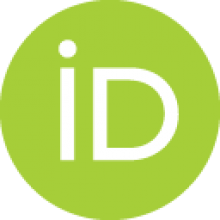}}}

\begin{document}

\title{Neutrino-Neutron Scattering Opacities in Supernova Matter} 

\author{Gang Guo\orcid{0000-0003-0859-3245}}
\email{guogang@cug.edu.cn} 
\affiliation{School of Mathematics and Physics, China University of Geosciences, Wuhan 430074, China}

\author{Gabriel Mart\'inez-Pinedo\orcid{0000-0002-3825-0131}}
\affiliation{GSI Helmholtzzentrum {f\"ur} Schwerionenforschung,
  Planckstra{\ss}e 1, D-64291 Darmstadt, Germany} 
\affiliation{Institut f{\"u}r Kernphysik (Theoriezentrum),
    Fachbereich Physik, Technische Universit{\"a}t Darmstadt,
    Schlossgartenstra{\ss}e 2, D-64289 Darmstadt, Germany} 

\author{Meng-Ru Wu\orcid{0000-0003-4960-8706}}
\affiliation{Institute of Physics, Academia Sinica, Taipei, 11529, Taiwan}
\affiliation{Institute of Astronomy and Astrophysics, Academia Sinica, Taipei, 10617, Taiwan}
\affiliation{Physics Division, National Center for Theoretical Sciences, Taipei 10617, Taiwan}

\date{\today}

\begin{abstract}
We compute the static density and spin structure factors in the long wavelength limit for pure neutron matter at subsaturation densities relevant to core-collapse supernovae within the Brueckner-Hartree-Fock (BHF) approach. The BHF results are reliable at high densities, extending beyond the 
validity of the virial expansion. Motivated by the similarities between the dilute neutron gas and a unitary gas, we propose a phenomenological approach to derive the static structures with finite momentum transfer as well as the dynamic ones with simple analytical expressions, based on the computed static structures in the long wavelength limit. We also compare the in-medium neutrino-neutron scattering cross sections using different structure factors. Our study 
emphasizes the importance of accurately computing the static structure factors theoretically  
and utilizing the full dynamic structure factors in core-collapse supernova simulations.  
\end{abstract} 

\maketitle

\section{Introduction}
Neutrino interactions with nucleons in hot and dense nuclear matter play an essential role in the dynamics and the associated synthesis of elements in core-collapse supernovae (CCSNe), as well as the mergers of neutron stars \cite{Burrows:2006,Janka:2007,Janka:2012,Burrows:2013,Burrows:2018,Radice:2020,Cowan:2019pkx,Fischer:2023}. As extensively investigated in earlier literature, the strong interaction between nucleons can have a significant impact on the neutrino-nucleon reaction rates in the nuclear medium. Such medium effects on neutrino rates due to nucleon interaction can be studied in the mean field approximation \cite{Reddy_1998,Vidana.Logoteta.Bombaci:2022} (see also \cite{Roberts:2017,Fischer:2020,Guo:2020} for studying the charged-current reactions incorporating higher-order weak interaction terms) and/or in the framework of random phase approximation (RPA) \cite{Reddy_1998,Burrows:1998cg,Burrows:1998ek,Reddy:1998hb,Oertel.Pascal.ea:2020}. While RPA calculations are model dependent, they have the advantage of being potentially consistent with many of the available equations of state (EoS) that are typically based on mean-field approaches.

At low density and/or high temperature, the virial expansion, using the measured scattering phase shifts, can provide a model-independent description of the EoS of nuclear matter \cite{Horowitz:2005zv,Horowitz:2005nd}, as well as the neutrino response, which is closely related to the nuclear EoS \cite{Horowitz:2006pj,Horowitz:2016gul}. 
The virial results are expanded in the fugacity of nucleons, $z=e^{\mu/T}$, with $\mu$ the non-relativistic chemical potentials. The virial approach is applicable for $z_{n,p} \ll 1$, which could be relevant to the typical conditions  encountered in the neutrinosphere in CCSNe with $\rho \approx 10^{11}$--$10^{13}$~g/cm$^{3}$ and $T \approx 5$--10 MeV. Given the non-linearity and complexity of neutrino transport in supernovae, it would be also desirable to have
reliable neutrino-nucleon reaction rates for a wider range of conditions (e.g., $z_{n,p}\lesssim$ a few) from microscopic calculations. Furthermore, the virial studies mentioned are limited to providing the static response functions or structure factors in the long wavelength limit, neglecting both energy and momentum transfer in neutrino-nucleon scattering (see, however, a recent study \cite{Bedaque_2018} for computing the dynamic structure factors in the virial expansion).

In this work, we present a microscopic calculation of the neutral-current neutrino-nucleon  scattering cross section using the Brueckner-Hartree-Fock (BHF) formalism (see, e.g., \cite{Baldo:2001,Nicotra:2006,Li:2010}), which is expected to be applicable to a broader range of conditions relevant to supernova matter.
To avoid the complexity arising from light cluster~\cite{Fischer.Typel.ea:2020} and pasta formation~\cite{Schuetrumpf.Martinez-Pinedo.Reinhard:2020}, we limit our discussion to pure neutron matter, as done in \cite{Horowitz:2006pj,Bedaque_2018}. Since 
$\beta$-equilibrium can be achieved
near the neutrinosphere with relatively high temperature, the fractions of protons and light cluster are typically much smaller than 10\% and neutrons contributes predominantly to the neutrino scattering rates. It should be also pointed out that the BHF method only provides the static structure factors in the long wavelength limit, obtained by taking the thermodynamic derivatives of the obtained EoS. Given that the dilute neutron gas shares similar properties as a unitary gas \cite{Schwenk:2005ka,Carlson:2012mh,Lin:2017,Horikoshi:2019ewa,Vidana:2021wna,Sammarruca:2023vmx}, owing to a large neutron-neutron scattering length, the well-studied structure factors of a unitary gas could offer insights into constructing the static structure factors for the dilute neutron gas with finite momentum transfer, as well the dynamic ones with both energy and momentum dependencies. We also compare in this paper the in-medium neutrino-neutron cross sections
using the obtained static and 
dynamic structure factors.

We organise this paper as follows. In Sec.~\ref{sec:S0}, we compare the static density and spin structure factors in the long wavelength limit using both the virial approach and the BHF scheme. Motivated by the similarity between a dilute neutron gas and a unitary gas, in Sec.~\ref{sec:Sq} we propose methods to obtain the static structure factors with finite momentum transfer, based on results
for a unitary gas as found in existing literature. The
derivation of dynamic structures using sum rules is detailed in Sec.~\ref{sec:sdyn}. In Sec.~\ref{sec:opa}, we compare the neutron-neutron scattering rates using different structure factors. We conclude this work in Sec.~\ref{sec:sum}. Throughout this work, we take the natural units with $\hbar = c = k_B =1$.

\section{static structure factors in the long wavelength limit, $S_{V,A}(0)$}   
\label{sec:S0}

\subsection{virial expansion} 

Below we present the basic formalism for the virial approach which can be applied to study the neutrino response of neutron matter at low densities and/or high temperatures. We refer to Refs.~\cite{Sedrakian.Roepke:1998,Horowitz:2005zv,Horowitz:2006pj,Roepke.Bastian.ea:2013} for additional details.
Analogous to the grand-canonical partition functions, thermodynamic quantities such as pressure ($P$), number density ($n$), and entropy ($S$) can be expressed as expansions in powers of the fugacity $z\equiv e^{\mu/T}$ with $T$ the temperature and $\mu$ the chemical potential excluding the nucleon rest mass~\cite{Horowitz:2005zv}. The expansion coefficients are linked to the so-called virial coefficients. For a dilute system with $z\lesssim 1/2$, keeping terms up to ${\cal O}(z^2)$ is generally sufficient. The related second virial coefficient is directly related to the two-body scattering phase shifts that are well measured experimentally.   

To study the spin response function, it is necessary to introduce $n_\sigma$, $\mu_\sigma$, and $z_\sigma=e^{\mu_\sigma/T}$ as the densities, chemical potentials, and fugacities of spin-up ($\sigma=\frac{1}{2}$) and spin-down ($\sigma=-\frac{1}{2}$) neutrons. The second virial coefficient and the second spin (axial) virial coefficient are given by
\begin{align}
b_n &= b_{n,1}+b_{n,0}, \\      
b_a &= b_{n,1}-b_{n,0},    
\end{align}
where $b_{n,1}$ and $b_{n,0}$ are the second virial coefficients for scattering of two incoming neutrons with same and opposite spin projections,
respectively. They are given by   
\begin{align}
b_{n,1}(T) &= \frac{2^{3/2}}{\pi T} \int dE e^{-E/T} \delta_1(E)-2^{-5/2},  \label{eq:bn1} \\
b_{n,0}(T) &= \frac{2^{3/2}}{\pi T} \int dE e^{-E/T} \delta_0(E), \label{eq:bn0}
\end{align}
where $E=k^2/m_n$ is the scattering energy with $m_n$ the neutron mass, $k=|\bm{p}-\bm{p}'|/2$ is the relative momentum between the two neutrons in the center-of-mass frame, and $\delta_{1,0}$ denote the total phase shift of partial waves: 
\begin{align}
\label{eq:phase1}
\delta_1(E) &= \sum_{L,J} \frac{2J+1}{3}\delta_L^{JS=1}(E), \\
\delta_0(E) &= \sum_{S,L,J} \frac{2J+1}{2(2S+1)} \delta_L^{JS}(E),  
\label{eq:phase0} 
\end{align}
with $L$, $S$, and $J$ the total orbital angular momentum, the total spin, and the total angular momentum of the neutron pair, respectively.

The scattering of neutrinos with nuclear matter can be characterized by the density (vector) dynamic structure factor $S_V(q, \omega)$ 
and the spin (axial) dynamic structure factor $S_A(q, \omega)$, where $q$ and $\omega$ represent the momentum and energy transferred from the neutrino to the nuclear medium. Under typical conditions around neutrinosphere, the scattering cross section could be approximated using the static structure factors \cite{Horowitz:2006pj}, which are defined as: 
\begin{align}
S_{V, A}(q) = \int_{-\infty}^{\infty} S_{V,A}(q, \omega) d\omega.  
\label{eq:sva_def}
\end{align}

For spin-unpolarised neutron matter, the static density structure factor in the long-wavelength limit is given by \cite{Horowitz:2006pj} 
\begin{align}
\label{eq:sv} 
S_V \equiv S_V(q=0) = \frac{z}{n} \left(\frac{\partial n}{\partial z}\right)_{T} = \frac{1+4b_n z}{1+2b_n z}.  
\end{align}
Similarly, the static spin structure factor of the unpolarised neutron matter is \cite{Horowitz:2006pj}
\begin{align}
\label{eq:sa} 
 S_A \equiv S_A(q=0) & = \frac{z_a}{n} \left(\frac{\partial n_a}{\partial z_a}\right)_{T,\,z_a=1} \nonumber \\
 &= 1 + \frac{2b_a z}{1+2b_n z}, 
\end{align}
where $n_a \equiv n_{\frac{1}{2}}-n_{-\frac{1}{2}}$ is the polarisation density, and $z_a\equiv \sqrt{z_{\frac{1}{2}}\Big/z_{-\frac{1}{2}}}$ is the axial fugacity.
In our numerical study, we derive the temperature-dependent virial coefficients from Eqs.~\eqref{eq:bn1} and \eqref{eq:bn0} considering the measured $T=1$ $np$ phase shifts up to $J=6$. For an approximate study, one can adopt values such as $b_n \approx 0.3$ and $b_a \approx -0.6$ \cite{Horowitz:2006pj,Bedaque_2018} under conditions relevant to SN matter, neglecting their relatively mild temperature dependencies.

\subsection{Brueckner-Hartree-Fock (BHF) scheme}
\label{sec:bhf}

The static structure factors can be derived from the EoS of neutron matter studied within the framework of the BHF formalism \cite{Bombaci:2005vi}. 
The neutron self-energy is given by  
\begin{align} 
\label{eq:self}
\Sigma_\sigma&(p, \varepsilon_\sigma(p)) =\frac{1}{2\pi} \sum_{\sigma'=-\frac{1}{2}}^{\frac{1}{2}} \int \frac{d^3p'}{(2\pi)^3} \sum_{Jll'SS_z}^{S+l\rm{~is~even}}
\Big(C^{SS_z}_{\frac{1}{2}\sigma\frac{1}{2}\sigma'}\Big)^2 
\nonumber \\ 
 \times & \sum_{Lm} (-1)^m C_{lmSS_z}^{J(m+S_z)} C_{l'mSS_z}^{J(m+S_z)} C_{lml'(-m)}^{L0} C_{l0l'0}^{L0} [ll'] P_L(\hat{\bm k}) \nonumber \\
 \times & 
 f_{\sigma'}(p') 
G_{ll'}^{JSS_z}(k, k; P, \Omega =  \varepsilon_\sigma(p)+\varepsilon_{\sigma'}(p')),
\end{align}
where $k = |\bm{k}| = \frac{1}{2}|\bm{p}-\bm{p'}|$ is the relative momentum between the neutron pair, $P=|\bm{P}|=|\bm{p}+\bm{p}'|$ is the total momentum, $[n]\equiv \sqrt{2n+1}$, $f_\sigma(p) = [1+e^{(\varepsilon_\sigma(p)-\mu_\sigma)/T}]^{-1}$ is the standard Fermi-Dirac distribution with $\varepsilon_\sigma(p)$ being the quasiparticle energy of the neutron, and $G_{ll}^{JSS_z}$ is the diagonal $G$-matrix element in partial wave basis with $S_z$ representing the spin projection. Note that in the above summation,
the isospin index has been omitted as it always equals 1 for pure neutron matter.
The $G$-matrix element can be obtained by solving the Bethe-Goldstone equation: 
\begin{align}
\label{eq:Gmatrix}
 G_{ll'}^{JSS_z}(k', k; P, \Omega) =& V_{ll'}^{JS}(k', k) + \sum_{l''} \frac{k''^2 dk''}{(2\pi)^3} V_{ll''}^{JS}(k', k'') \nonumber \\
 & \times  \bar g_{II}^{S_z}(P, \Omega, k'')  G_{l''l'}^{JSS_z}(k'',k), 
\end{align}   
where $V^{JS}_{ll'}$ is the neutron-neutron potential in partial wave basis, and $\bar g_{II}^{S_z}(P, \Omega, k)$ is the angle-averaged two-neutron propagator. In this work, we employ the two-body potential of chiral effective field theory ($\chi$EFT) up to N$^4$LO from \cite{Entem:2017gor}. In the quasiparticle approximation, $\bar g_{II}^{S_z}$ is given by 
\begin{align}
\bar {g}_{II}^{S_z}(P, \Omega, k) = \biggl\langle \frac{[1-f_\sigma( p_1)][1-f_{\sigma'}(p_2)]}{\Omega - \varepsilon_\sigma(p_1) - \varepsilon_{\sigma'}(p_2)+ i\eta} \biggr\rangle_{\theta}.    
\label{eq:gII}  
\end{align}
Here, $\sigma$ and $\sigma'$ represent any set of spin alignments satisfying $\sigma+\sigma'=S_z$, $p_{1,2}=|\bm{p}_{1,2}| = |{1\over 2}\bm{P} \pm \bm{k}|$ are the momenta of the two intermediate neutrons in the propagator, and $\theta$ is the angle between the total momentum $\bm{P}$ and the relative momentum $\bm{k}$.

\begin{figure}[htbp]  
\centering
\includegraphics[width=0.49\textwidth]{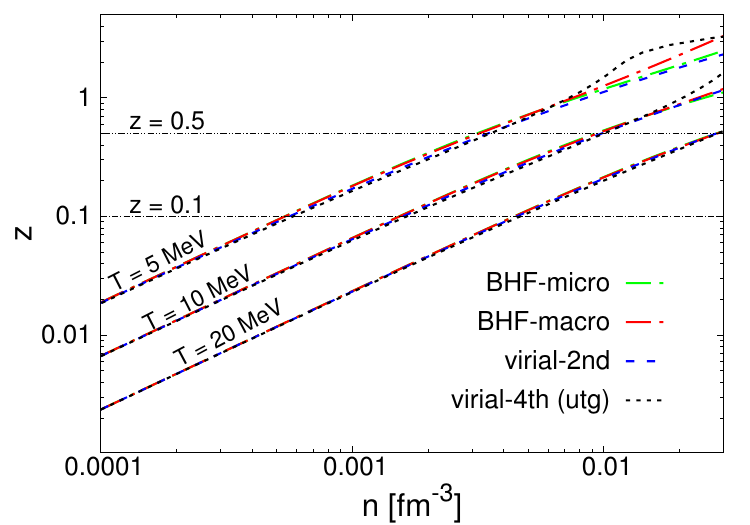}
\caption{Fugacities for unpolarised neutron matter as functions of neutron number density at $T=5$, 10, and 20 MeV from the virial expansion up to ${\cal O}(z^2)$, the microscopic and the macroscopic calculations within the BHF scheme, and the virial expansion up to ${\cal O}(z^4)$ for a unitary gas. The horizontal lines correspond to $z=0.1$ and $z=0.5$.} 
\label{fig:nz}  
\end{figure}

\begin{figure*}[htbp]  
\centering
\includegraphics[width=\textwidth]{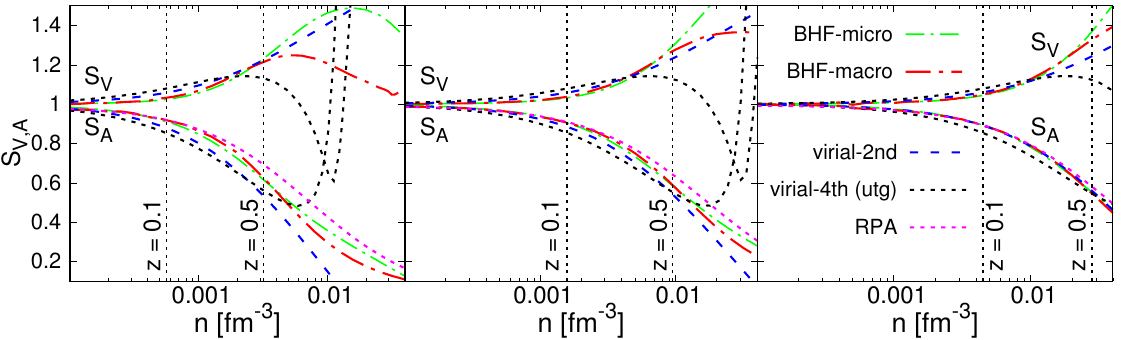}
\caption{The static structure factors, $S_{V, A}\equiv S_{V,A}(0)$, as functions of density at $T=5$ MeV (left), 10 MeV (middle), and 20 MeV (right) from the virial expansion up to ${\cal O}(z^2)$, the microscopic and the macroscopic calculations within the BHF scheme, respectively. For comparison, the RPA results of $S_A$ from \cite{Burrows:1998cg}
and the virial results up to ${\cal O}(z^4)$ for a unitary gas from \cite{Lin:2017} are also shown.} 
\label{fig:sva} 
\end{figure*}

The self-consistent solution for the BHF quasiparticle spectrum and $z_\sigma$ can be obtained iteratively with Eqs.~(\ref{eq:self}-\ref{eq:gII}), considering the
on-shell condition
\begin{align}
& \varepsilon_\sigma(p) = \frac{p^2}{2m_n} + {\rm Re}\Sigma_\sigma(p, \varepsilon_\sigma(p)), \label{eq:eps}
\end{align}
and the number density $n_\sigma = \int \frac{d^3p}{(2\pi)^3} f_\sigma(p)$.

For unpolarised neutron matter, Eq.~(\ref{eq:self}) simplifies to  
\begin{align} 
\label{eq:self_sym} 
\Sigma(p, \varepsilon(p)) =& \Sigma_\sigma(p, \varepsilon_\sigma(p)) = \frac{1}{2} \sum_\sigma \Sigma_\sigma(p, \varepsilon_\sigma(p)) \nonumber \\
=& \frac{1}{4\pi}\int \frac{d^3p'}{(2\pi)^3} \sum_{JLS}^{S+L\rm{~is~even}}
(2J+1) f(p') 
\nonumber \\ 
 & \times  
G_{LL}^{JS}(k, k; P, \Omega =  \varepsilon(p)+\varepsilon(p')).
\end{align}

The static structure factors, in the long wavelength limit, can be obtained from the derivative of neutron density with respect to fugacities $z=z_{1/2}=z_{-1/2}$ and $z_a$ for unpolarised neutron matter [see the first expressions in Eqs.~(\ref{eq:sv}) and (\ref{eq:sa})]. The quasiparticle chemical potentials $\mu_\sigma$ and fugacities $z_\sigma\equiv e^{\mu_\sigma/T}$ can be straightforwardly obtained from Eqs.~(\ref{eq:self}-\ref{eq:eps}) through iteration. These are commonly referred to as the microscopic (axial) chemical potentials and (axial) fugacities in the literature. Apart from these microscopically computed ones for quasiparticles, the chemical potentials can also be derived by taking the derivative of the free energy with respect to (polarisation) densities \cite{Rios:2008fz}:
\begin{align}
\label{eq:mu_mac} 
& \tilde \mu = T\ln(\tilde z) = \left(\frac{\partial f}{\partial n}\right)_T, \\
& \tilde \mu_a = T \ln(\tilde z_a) = \left(\frac{\partial f}{\partial n_a} \right)_{T, n, n_a=0}, 
\label{eq:mua_mac}
\end{align}
with $f$ being the free energy density. To distinguish them from the microscopic ones, the (axial) chemical potentials or (axial) fugacities [Eqs.~\eqref{eq:mu_mac} and \eqref{eq:mua_mac}] derived using thermodynamic relations are termed the macroscopic ones. The free energy $f$ can be obtained from the internal energy density $\epsilon=E/V$ and the entropy density $s=S/V$ as follows:   
\begin{align} 
& f = \epsilon - T s, \label{eq:f}\\
& \epsilon = \sum_\sigma\int \frac{d^3p}{(2\pi)^3} \left[ \frac{p^2}{2m_n} + \frac{1}{2} {\rm Re}\Sigma_\sigma(p, \varepsilon_\sigma(p)) \right] f_\sigma(p),  \label{eq:eps} \\
& s = -\sum_\sigma\int \frac{d^3p}{(2\pi)^3} \left[ f_\sigma(p) \ln\Big(f_\sigma(p)\Big) \right. \nonumber \\
& \qquad\qquad\qquad + \left. \Big(1- f_\sigma(p)\Big)\ln\Big(1- f_\sigma(p)\Big) \right].\label{eq:s}
\end{align} 
Note that in Eqs.~(\ref{eq:f}-\ref{eq:s}), the neutron quasiparticle energies as well as the microscopic chemical potentials are used.

It is important to note that the BHF approach at finite temperature can not give thermodynamically consistent chemical potentials (i.e., $\mu \ne \tilde \mu$, $\mu_a \ne \tilde\mu_a$) at densities close to or higher than the nuclear saturation density, due to the absence of hole-hole contributions \cite{Rios:2008fz,Baldo:1999}. The hole-hole correlations can be taken into account within the self-consistent Green's function (SCGF) approach \cite{Muther:2000,Frick:2003,Dickhoff:2004,Carbone:2013,Carbone:2013b}, which, however, is more involved to solve numerically than the BHF approach. Ref.~\cite{Rios:2008fz} performed a systematic study of pure neutron matter at finite temperature within the SCGF approach using the CD Bonn \cite{Machleidt:2000ge} and the Argonne V18 \cite{Wiringa:1994wb} potentials, demonstrating consistent microscopic and macroscopic chemical potentials for each potential. In addition, the macroscopic chemical potentials from the BHF approach were found to agree well with those from the SCGF approach at subsaturation densities \cite{Rios:2008fz}. However, the microscopic ones within BHF are reliable only at low densities, tending to underestimate the chemical potentials at $n\gtrsim 0.01~\rm{fm}^{-3}$~\cite{Rios:2008fz}.

\subsection{results of $S_{V,A}(0)$}

Figure~\ref{fig:nz} compares the fugacities for spin-symmetric neutron matter as functions of density at $T=5$, 10, and 20 MeV from the virial expansion and the BHF microscopic [Eqs.~(\ref{eq:self}-\ref{eq:eps})] and the macroscopic [Eq.~\eqref{eq:mu_mac}] calculations. As shown in the figure, the three different approaches lead to close results at low densities. Note that at the subsaturation densities encountered in supernova conditions, the fugacity can be as high as a few for $T=5$ MeV, while it typically stays below 1 for $T=20$ MeV. Considering the virial expansion is not valid above $z\approx 1$, our BHF results start to differ from the virial ones at high densities. The hole-hole correlation, which is neglected in the BHF microscopic calculations, tends to add a repulsive contribution to the neutron quasiparticle energy. However, such correlation effect can be properly taken into account in the BHF macroscopic calculation \cite{Rios:2008fz}, leading to a larger chemical potential or fugacity compared to the microscopic results within BHF. For comparison, Fig.~\ref{fig:nz} also shows the virial results up to fourth order for a unitary gas~\cite{Lin:2017}. The axial fugacity, $z_a$, can be computed similarly.   
       
The static vector and axial vector structure factors $S_{V,A}$ in the long wavelength limit can be obtained by taking the derivatives of the fugacity and the axial fugacity with respect to $n$ and $n_a$ [Eqs.
\eqref{eq:sv} and \eqref{eq:sa}], respectively. Depending on whether the microscopic or the macroscopic fugacities are utilised, one can obtain the corresponding microscopic and macroscopic $S_{V,A}$. As shown in Fig.~\ref{fig:sva}, the structure factors from different approaches are close to each other at low densities. The virial results for a unitary gas up to ${\cal O}(z^4)$ are also shown.
With $b_n \approx 0.3$ and $b_a \approx -0.6$, the virial expansion up to ${\cal O}(z^2)$ gives rise to an increase of $S_V$ and a decrease of $S_A$ with density [Eqs.~(\ref{eq:sv}) and (\ref{eq:sa})], as predicted by the BHF approach. As the vertex correction diagrams have not been taken into account, the BHF tends to underestimate the medium effects at low densities, resulting in smaller $S_V$ and larger $S_A$ compared to the virial results. The static structure factors from both methods agree well at intermediate values of $z$. At $z\gtrsim 0.5$, the virial expansion starts to fail and could even produce a negative $S_A$ at $z\gtrsim 1$. In this regime ($z\gtrsim 0.5$), the BHF results should be considered more reliable. For practical application, we suggest using the viral results of the neutron gas for $z\le 0.25$ and the BHF results for $z\ge 0.5$, and performing a numerical interpolation for the intermediate regime. We also show the RPA results for $S_A(0)$ based on \cite{Burrows:1998cg}. We find that the RPA studies tend to overestimate $S_A(0)$ at intermediate and high densities, compared to our BHF calculations. Note that, in contrast to the virial and the BHF calculations, the RPA results for $S_V(0)$ are smaller than 1 and are not shown.

\begin{table*}[htbp] 
  \centering 
  \caption{Fitting coefficients for fugacity $z$ and static structure factors $S_{V,A}(0)$ as functions of $T$ and $n$, see Eqs.~\eqref{eq:fitS} and \eqref{eq:fitz}. Note that we use the notation that $a(b)\equiv a \times 10^b$ for the numerical values shown in the table. The fit is valid for $5\le T \le 30$ MeV and $10^{-4} \le n \le 0.063$~fm$^{-3}$. \label{tab:fit_coeff}}
\begin{ruledtabular}
\renewcommand{\arraystretch}{1.5}
\begin{tabular}{ccccccccc}    
  & $c_1^{z,V,A}$ & $c_2^{z,V,A}$ & $c_3^{z,V,A}$ & $c_4^{z,V,A}$ & $c_5^{z,V,A}$ & $c_6^{z,V,A}$ & $c_7^{z,V,A}$ & $c_8^{z,V,A}$  \\    
\hline
$z$ &  8.373($-$1) & $-$2.959($-$3) & $-$2.851($-$1) & 9.174($-$3) & 5.156($-$2) & $-$1.001($-$2) & $-$5.276($-$3) & 1.480($-$3) \\   
$S_{V}(0)$ & 1.056 & $-$2.660($-$2) & $-$1.011 & $4.185($-$2)$ & 1.704($-$1) & $-$1.047($-$2) & $-$1.319($-$2) & 1.036($-$3)  \\
$S_{A}(0)$ &  $-$1.135 & $-$1.604($-$3) & 7.773($-$1) & $-$2.669($-$3) & $-$9.112($-$2) & 2.724($-$3) & 6.354($-$3) & $-$4.824($-$4)  \\
\end{tabular}
\end{ruledtabular}
\end{table*}

As the hole-hole contribution is negligible at low $z$, the BHF microscopic and macroscopic calculations lead to similar structure factors. At higher $z$, however, the structure factors $S_{V,A}$ obtained from the macroscopic chemical potential should be more reliable compared to the microscopic ones. Note that the macroscopic $S_V$ can be smaller than the microscopic ones by 0.4 in magnitude, while the difference between the microscopic and macroscopic $S_A$ is $\lesssim$ 0.05 for the conditions considered. 

In the following we provide fitting functions to the computed structure factors covering a wide range of temperature and density conditions. Temperature, $T$, ranges from 5 to 30 MeV and baryon density, $n$, from $10^{-4}$ to 0.063 fm$^{-3}$.  The largest value of $z\approx 7$ is obtained for $T=5$ MeV and $n=0.063$ fm$^{-3}$. We take the virial results for $z\le 0.25$ and the BHF results for $z \ge 0.5$. For intermediate $z$, we employ a numerical interpolation.
With this procedure, we fit the derived $S_{V,A}(0)$ as functions of $n$ and $T$ with the form suggested in Ref.~\cite{Lin:2017}:
\begin{align}
S_{x}(0) = 1 & + (c_1^x + c_2^x T) \eta^{3/2} + (c_3^x + c_4^x T) \eta^2 \nonumber \\
& + (c_5^x + c_6^x T) \eta^3 + (c_7^x + c_8^x T) \eta^4,
\label{eq:fitS}   
\end{align}
where $x$ denotes $V$ or $A$, and $\eta \equiv T_F/T$ with $T_F=(3\pi^2 n)^{2/3}/(2m_n)$ the Fermi temperature. The temperature $T$ in the expressions is in unit of MeV. Similar to the static structure factors, we also provide a fit for $z$ with the form 
\begin{align}
z = & (c_1^z + c_2^z T) \eta^{3/2} + (c_3^z + c_4^z T) \eta^2 \nonumber \\ + & (c_5^z + c_6^z T) \eta^3 + (c_7^z + c_8^z T) \eta^4. \label{eq:fitz}
\end{align}
All the fitting coefficients are listed in Table.~\ref{tab:fit_coeff}. The fits are accurate within 2\% for most of the conditions explored and become worse at $z\gtrsim 5$ where the error can be as large as 5\%.

\section{Static structure factors with finite momentum transfer, $S_{V,A}(q)$}
\label{sec:Sq} 

In this section, we present a phenomenological approach to obtain the static structure factors, $S_{V,A}(q)$, with finite momentum transfer $q$. With a large scattering length, we assume that the dilute neutron gas exhibits similar properties of structure factors as a unitary gas. Consequently, we choose to use the extensively studied structure factors of a unitary gas to describe the $S_{V,A}(q)$ of a neutron gas.

Firstly, we assume that in the limit of large $q$,      
\begin{align}
S_V(q \gg k_F) \approx 1 + {I k_F \over 4q}, \label{eq:Sv_large}\\
S_A(q \gg k_F) \approx 1 - {I k_F \over 4q}, \label{eq:Sa_large}
\end{align}
where $k_F=(3\pi^2n)^{1/3}$ is the Fermi momentum, and $I$ is the Tan contact\footnote{Note that here $I$ denotes the dimensionless Tan contact \cite{Liu:2013}.} accounting for the short-distance behaviour of the pair correlation function. The static structure factors of neutron matter have recently been computed using a lattice formulation at specific conditions relevant to supernova matter  \cite{Alexandru:2020zti}, confirming the asymptotic behaviours shown in Eqs.~\eqref{eq:Sv_large} and \eqref{eq:Sa_large}. Moreover, the Tan contact of the neutron gas extracted from the calculated static structure factors is close to those of a unitary gas. In this work, we adopt a similar Tan contact to a unitary gas for the dilute neutron gas as explained below. The Tan contact of a unitary gas only depends on the dimensionless parameter $T/T_F$.
At high $T$ or low $z$, the virial expansion of the Tan contact up to ${\cal O}(z^3)$ is available for a unitary gas \cite{Liu:2013}. For simplicity, we take the same virial expansion coefficients to obtain the Tan contact of the neutron gas at low $z$. Experimental measurements of the Tan contact at low temperatures ($T \gtrsim 0.1 T_F$) have been conducted using Bragg spectroscopy of cold atomic Fermi gases and the unitary gas, albeit with relatively large uncertainties \cite{Sagi:2012,Carcy:2019,Mukherjee:2019}. To match the measured Tan contact at low temperatures, we find that a linear extrapolation of $I$ in $T/T_F$ derived from the virial expansion, starting from $T/T_F\approx 1$ and extending to low $T$, is a suitable choice (see, e.g., Fig.~3 of \cite{Alexandru:2020zti}). Based on the above arguments, we take the following expressions for the Tan contact of a neutron gas: 
\begin{align}
I = \left \{ 
\begin{array}{cc}
3\pi^2 \Big({T \over T_F}\Big)^2 ({z^2 \over \pi} - 0.14 z^3), & z \le 0.5\;,  \\
2.22 + 0.4(1.1 - {T\over T_F}), & z >0.5\;.
\end{array}
\right.
\end{align}
Note that for a neutron gas $T/T_F \approx 1.1$ at $z=0.5$. 

\begin{figure}[htbp]  
\centering
\includegraphics[width=0.49\textwidth]{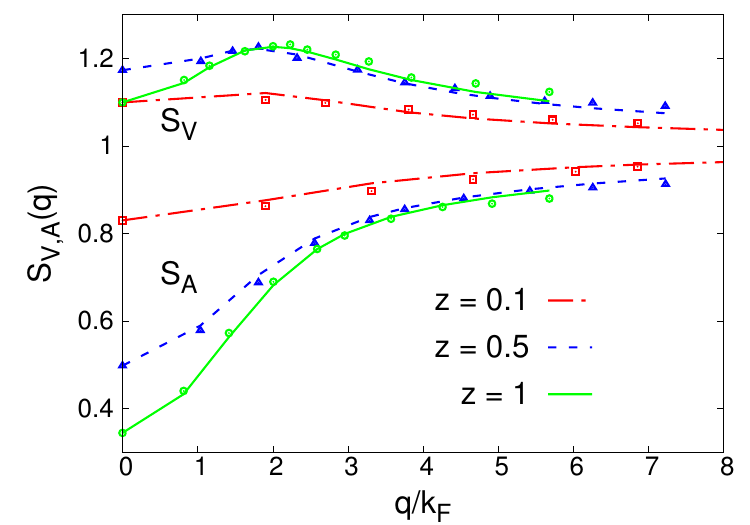}
\caption{
Fits to $S_{V,A}(q)$ for a unitary gas under supernova conditions computed from a lattice formulation \cite{Alexandru:2019gmp} using the expressions given by Eqs.~\eqref{eq:SVq} and \eqref{eq:SAq}.} 
\label{fig:Sq_fit}  
\end{figure}

The static structure factors, $S_{V,A}(q)$, of a unitary gas have also been computed with the lattice formulation for a wide range of conditions relevant to supernova matter beyond the validity of the virial expansion \cite{Alexandru:2019gmp}. We suggest that the obtained $S_{V,A}(q)$ can be well fitted with the expressions (see Fig.~\ref{fig:Sq_fit})
\begin{align}
S_{V}(q) =  1 & + \Big[S_{V}(0) - 1 - {a_V I \over 4}\Big]  e^{-b_V x_q^2} \nonumber \\
& + {I \over 4 x_q} \tanh{(a_V x_q)}, \label{eq:SVq} \\
S_{A}(q) = 1 & + \Big[S_{A}(0) - 1 + {a_A I\over 4}\Big]  e^{-b_A x_q^2} \nonumber \\
& - {I \over 4x_q} \tanh{(a_A x_q)}, \label{eq:SAq}
\end{align}
where $x_q \equiv q/k_F$, and the coefficients are $a_V=1.028$, $b_V=0.555$, $a_A=0.637$, $b_A=0.346$.
It can be readily verified that in the limit $q/k_F \gg 1$ the static structure factors adhere to the asymptotic behaviours outlined in Eqs.~\eqref{eq:Sv_large} and \eqref{eq:Sa_large}.  We also note that Eqs.~\eqref{eq:SVq} and \eqref{eq:SAq} are formulated to be even functions of $q$, ensuring that $dS_{V,A}(q)/dq$ vanishes at $q=0$. This is consistent with the fact that $S_{V,A}(q, \omega)$ represent the Fourier transforms of the corresponding pair correlation functions. 

We employ Eqs.~\eqref{eq:SVq} and \eqref{eq:SAq} along with the $S_{V,A}(0)$ obtained from either our BHF calculation or the virial expansion to determine the $S_{V,A}(q)$ for the dilute neutron gas. It should be stressed, however, that the neutron gas, with a finite effective range, could be different from a unitary gas \cite{Vidana:2021wna}. Despite that, the slowly varying nature of $S_{V,A}$ at small $q$ and the asymptotic behaviour at $q \gtrsim 2k_F$ have already imposed significant constraints on $S_{V,A}(q)$. Therefore, the method we propose for constructing the $S_{V,A}(q)$ of the neutron gas should, at the very least, capture qualitative aspects and, in general, allow for quantitative results.

\section{dynamic structure factors, $S_{V,A}(q, \omega)$} 
\label{sec:sdyn}

Similarly to the static structure factors, the dynamic structure factors of a strongly interacting Fermi gas, especially a unitary gas, could provide valuable insights into understanding the qualitative behaviours of the dynamic structure factors of the neutron gas. For ultracold atomic gases with tunable interaction strength spanning the Bose-Einstein condensate (BEC)-Bardeen-Cooper-Schrieffer (BCS) crossover, both the static and the dynamic structures have been widely measured using 
Bragg spectroscopy. At high temperatures, $S_{V,A}(q, \omega)$ of a strongly correlated Fermi gas  with a short-range attractive $S$-wave potential can be virially computed up to ${\cal O}(z^2)$ \cite{Hu_2010,Liu:2013}, consistent with those extracted from experiments. As anticipated, the dynamic structure factor peaks at $\omega = \omega_q/2 = q^2/(4m)$ with a width of $\sqrt{\omega_q T}$ in the BEC limit, and peaks at $\omega = \omega_q = q^2/(2m)$ with a width of $\sqrt{2 \omega_q T}$ in the BCS limit, since the associated underlying constituents are bound molecules with masses of $2m$ and free atoms with masses of $m$, respectively. In the intermediate regime, including the unitary limit, a two-peak structure emerges from both the molecular and the atomic responses. It has also been observed that the peaks in dynamic structure factors can be well described by Gaussian functions across a wide range of interaction strengths \cite{Kuhnle_2011,Liu:2013}. 

Motivated by the above discussions for strongly interacting Fermi gas, we propose that the dynamic structure factors, $S_{V,A}(q, \omega)$, of the neutron gas can be quantitatively represented by the form     
\begin{align}
S_{V,A}(q, \omega) = A_{V,A}~e^{-{(\omega-\omega_q)^2 \over 4 \omega_q T}} + B_{V,A}~e^{-{(\omega-{1\over 2}\omega_q)^2 \over 2\omega_q T}},\label{eq:Sdyn_fit} 
\end{align}
with $\omega_q \equiv q^2/(2m_n)$. The coefficients $A_{V,A}$ and $B_{V,A}$ can be uniquely determined from the $f$-sum rules
\begin{align}
& \int_{-\infty}^{+\infty} d\omega \omega S_{V,A}(q, \omega) \nonumber \\
= & \int_{0}^{+\infty} d\omega (1-e^{-\omega/T}) \omega S_{V,A}(q, \omega)={q^2\over 2m_n},\label{eq:fsum}
\end{align}
together with the definitions of $S_{V,A}(q)$ with
\begin{align}
&\int_{-\infty}^{+\infty} d\omega S_{V,A}(q, \omega) \nonumber \\
= &\int_{0}^{+\infty} d\omega (1+e^{-\omega/T})S_{V,A}(q, \omega) = S_{V,A}(q).\label{eq:Sq_def}
\end{align}
Note that we have used the relation $S(q, -\omega)=e^{-\omega/T}S(q, \omega)$ in deriving Eqs.~\eqref{eq:fsum} and \eqref{eq:Sq_def}. It should be also pointed out that for $S_A$, Eq.~(\ref{eq:fsum}) is not exact as the nucleon spin is not conserved by nuclear interactions \cite{Sigl:1995ac,Bedaque_2018}. Here, we simply assume that the correction to Eq.~\eqref{eq:fsum} is relatively minor considering that the neutron interaction is dominated by the $^1S_0$ component \cite{Bedaque_2018}.
With the special forms given in Eq.~\eqref{eq:Sdyn_fit}, both integrals in Eqs.~\eqref{eq:fsum} and \eqref{eq:Sq_def} can be analytically solved, leading to 
\begin{align}
& A_{V,A} = [2-S_{V,A}(q)]\big/\sqrt{4\pi \omega_q T}, \label{eq:A_VA}\\
& B_{V,A} = \sqrt{2}[1-S_{V,A}(q)]/\sqrt{\pi \omega_q T}. \label{eq:B_VA}
\end{align}

\begin{figure}[htbp]  
\centering
\includegraphics[width=0.49\textwidth]{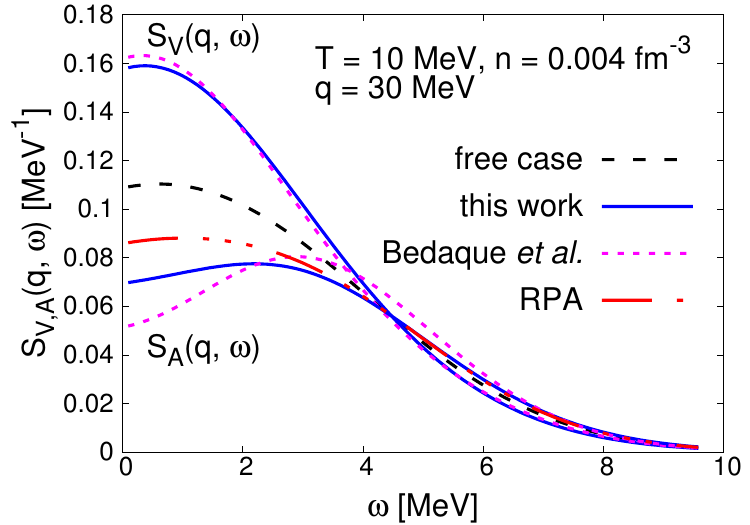}
\caption{Dynamic structure factors obtained from Eq.~\eqref{eq:Sdyn_fit} at $T=10$ MeV, $n=0.004$~fm$^{-3}$, and $q=30$ MeV. The dynamic structure factors from \cite{Bedaque_2018}, the $S_A(q, \omega)$ from RPA calculations \cite{Burrows:1998cg}, and the ones for a non-interacting neutron gas are also shown.} 
\label{fig:Sdyn}  
\end{figure} 

With $S_{V,A}(0)$ derived either from the BHF calculations or from the virial expansion, the $S_{V,A}(q, \omega)$ can be easily obtained by using Eqs.~\eqref{eq:SVq}, \eqref{eq:SAq}, and \eqref{eq:Sdyn_fit}. Figure~\ref{fig:Sdyn} presents our calculated $S_{V,A}(q, \omega)$ with $q=30$ MeV as function of $\omega$ at $T=10$ MeV and $n=0.004$ fm$^{-3}$ (corresponding to $z\approx 0.23$), where we have used $S_{V}(0)\approx 1.13$ and $S_A(0)\approx 0.75$ from the virial expansion. Compared to the non-interacting case \cite{Reddy_1998}, the nucleon interaction enhances $S_V(q, \omega)$ and suppresses $S_A(q, \omega)$ significantly at $\omega \lesssim {\rm max}[\omega_q, \sqrt{\omega_q T}]$. The dynamic structure factors computed from the virial expansion up to ${\cal O}(z^2)$ using a pseudopotential are also shown \cite{Bedaque_2018}.
Note that we have used the fitting formulae of \cite{Bedaque_2018} to produce their $S_{V,A}(q, \omega)$. We find that our $S_{V}(q, \omega)$ agrees well with that of \cite{Bedaque_2018}, while for $S_A(q, \omega)$, slightly larger differences occur. Note that our $S_A(q, \omega)$ is normalized to a slightly larger $S_A(q)\approx 0.75$ than that of \cite{Bedaque_2018} with $S_A(q) \approx 0.74$ at $q=30$ MeV. Using the same $S_A(q)$ of \cite{Bedaque_2018}, the derived $S_A(q, \omega)$ based on Eqs.~\eqref{eq:Sdyn_fit}, \eqref{eq:A_VA}, and \eqref{eq:B_VA} gets closer to the one of \cite{Bedaque_2018}. At low densities as considered in Fig.~\ref{fig:Sdyn}, the RPA results for $S_A(q, \omega)$ from \cite{Burrows:1998cg} are higher than our $S_A(q, \omega)$ at low $\omega$, which is consistent with the observations for $S_A(0)$ shown in Fig.~\ref{fig:sva}.

\section{neutrino-neutron scattering cross section}
\label{sec:opa}

In this section, we compare the neutral-current scattering cross sections of neutrinos with neutrons, using the different structure factors derived above. The double differential cross section is related to the full dynamic structure factors as 
\begin{align}
{d\sigma(E_\nu) \over d\omega d\cos\theta} = &~ {G_F^2\over 8\pi} (E_\nu-\omega)^2 [(1+\cos\theta) S_V(q, \omega) \nonumber \\
& + g_a^2 (3-\cos\theta) S_A(q, \omega)],  \label{eq:dsigdwdmu}
\end{align}
where $E_\nu$ is the incoming neutrino energy, $\theta$ is the scattering angle, $G_F$ is the Fermi constant, and $g_a \approx 1.26$ is the axial coupling constant. Integrating over the kinetically allowed energy transfer $\omega$ and/or $\cos\theta$, the single differential cross sections $d\sigma(E_\nu)/d\cos\theta$,and $d\sigma(E_\nu)/d\omega$, and the total cross section $\sigma(E_\nu)$ can be derived. Note that the associated integration limits are $-q\le \omega \le {\rm min}[2E_\nu-q, q]$ and $-1\le \cos\theta \le 1$. The opacities for neutrino-neutron scattering can be obtained by multiplying the scattering cross sections with the neutron number density.

\begin{figure*}[htbp]
\centering  
\includegraphics[width=\textwidth]{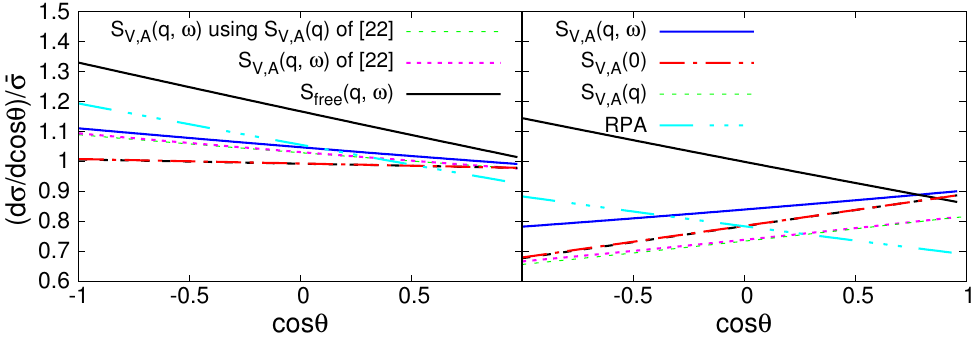} \\
\includegraphics[width=\textwidth]{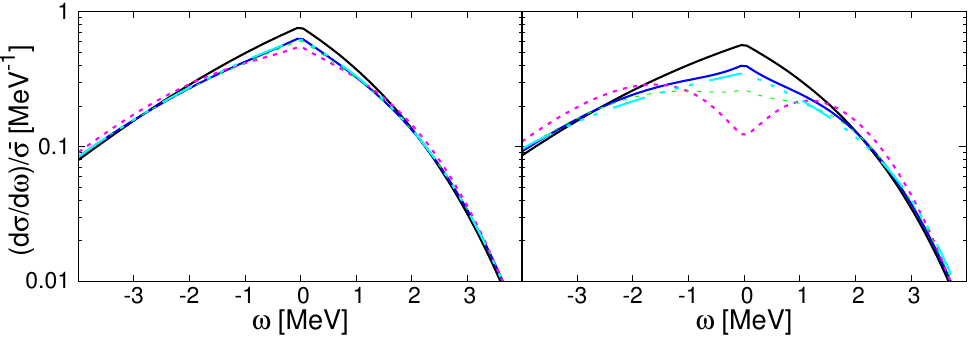} \\
\includegraphics[width=\textwidth]{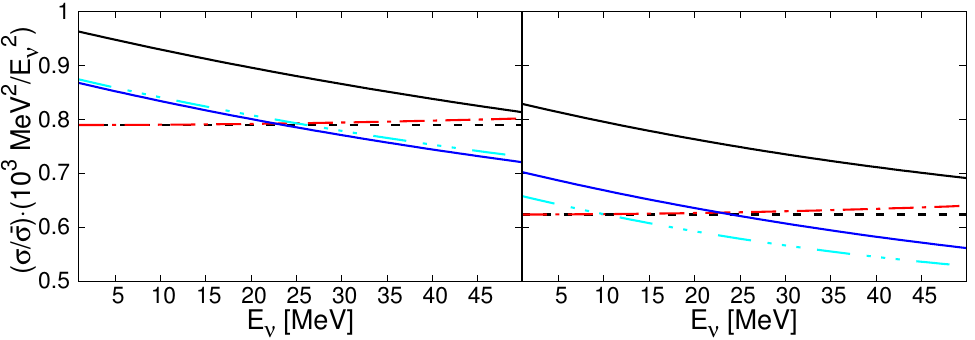}
\caption{Differential cross sections, $(d\sigma/d\cos\theta)/\bar\sigma$ (upper) and $(d\sigma/d\omega)/\bar\sigma$ (middle), and total cross sections $[\sigma(E_\nu)/\bar\sigma]\cdot (10^3$ MeV$^2/E_\nu^2)$ (lower) using different structure factors for two selected conditions: $T=10$ MeV and $n=0.004$~fm$^{-3}$ (left), and $T=10$ MeV and $n=0.0126$~fm$^{-3}$ (right). Note that all the cross sections shown are normalized by $\bar\sigma = 10^{-42}$~cm$^2$. For the differential cross sections (upper and middle), we have chosen $E_\nu = 10$ MeV. For the total cross section (lower), an additional factor of $10^3$ MeV$^2/E_\nu^2$ has been added to eliminate the apparent $E_\nu^2$-dependence.}
\label{fig:Sig}  
\end{figure*}

If only the static structure factors $S_{V,A}(q)$ or even just $S_{V,A}(0)$ are available, $d\sigma(E_\nu)/d\cos\theta$,  in the so-called elastic limit, could be approximated by  \cite{Bruenn1985,Lin:2017,Bedaque_2018}
\begin{align}
{d\sigma(E_\nu) \over d\cos\theta} \approx {G_F^2\over 8\pi} E_\nu^2 & [(1+\cos\theta) S_V(q) \nonumber \\
 +& g_a^2 (3-\cos\theta) S_A(q)]   \label{eq:dsigdmu0} \\
\approx {G_F^2\over 8\pi} E_\nu^2 & [(1+\cos\theta) S_V(0) \nonumber \\
 + & g_a^2 (3-\cos\theta) S_A(0)].\label{eq:dsigdmu} 
\end{align} 
As previously mentioned, $S_{V,A}(q, \omega)$ peaks around $\omega \approx q^2/(2m_n)$ and exhibits a width $\approx \sqrt{q^2T/m_n}$ in the energy distribution for a given $q$.
Therefore, it is imperative to verify the validity of the elastic approximation, particularly for high temperatures and/or $E_\nu$.

We compare in Fig.~\ref{fig:Sig}
the differential and the total cross sections using different static and dynamic structure factors obtained for two selected conditions: $T=10$ MeV and $n=0.004~{\rm fm^{-3}}$ with $z\approx 0.23$, and $T=10$ MeV and $n=0.0126~{\rm fm^{-3}}$ with $z\approx 0.62$. For the first condition, we use the virial results to derive the structure factors, while for the second one, we rely on the BHF results.
For comparison, we also show the results using the dynamic structure factor for a non-interacting neutron gas with $S_{\rm V, free}(q, \omega) = S_{\rm A, free}(q, \omega) = S_{\rm free}(q, \omega)$ [see Eq. (21) of \cite{Reddy_1998}\footnote{Note that, with a different notation, $S_{V,A}(q, \omega)$ in \cite{Reddy_1998} should be divided by $2\pi n$ for comparisons with the ones in this work.}].
As expected, the in-medium cross sections are generally suppressed by 10\%--30\% compared to the ones for a non-interacting neutron gas, primarily due to the significant reduction in the spin structure factor resulting from neutron-neutron interactions (see Fig.~\ref{fig:sva}).   

For backward scattering (i.e., $\cos\theta = -1$), only the spin structure factor contributes [see Eq.~\eqref{eq:dsigdwdmu}]. In this case, the reduction in $d\sigma/d\cos\theta$ with $E_\nu=10$ MeV compared to the non-interacting case is maximal (upper panels of Fig.~\ref{fig:Sig}) as the spin structure factor is suppressed. 
As $\cos\theta$ increases, the density response becomes more relevant, leading to closer $d\sigma/d\cos\theta$ from various treatments to that for the non-interacting case. With the dynamic spin structure factors being more suppressed at higher densities, the density structure factor could be dominant. As a consequence, $d\sigma/d\cos\theta$ will increase with $\cos\theta$ [see Eq.~\eqref{eq:dsigdwdmu} and the upper right panel of Fig.~\ref{fig:Sig}]. The differential cross sections as well as the total cross sections using $S_{V,A}(0)$ and $S_{V,A}(q)$ are very similar. The reason is that the typical momentum transfer $q$ is of order $E_\nu \ll k_F \approx 100$ MeV, and the related $S_{V,A}(q) \approx S_{V,A}(0)$ since $S_{V,A}(q)$ vary slowly at small $q$. As $E_\nu$ (and correspondingly, the related $q$) becomes comparable to $k_F$, the cross section using $S_{V,A}(q)$ becomes slightly large than that based on $S_{V,A}(0)$, as shown in the lower panels of Fig.~\ref{fig:Sig}.

The use of static structure factors instead of the dynamic one has some apparent limitations. Firstly, the energy of the outgoing neutrino, which determines the scattering phase space and consequently impacts the scattering cross section [see the $(E_\nu-\omega)^2$ factor in Eq.~\eqref{eq:dsigdwdmu}], can not be correctly computed in the elastic approximation [see the $E_\nu^2$ factor in Eqs.~\eqref{eq:dsigdmu0} and \eqref{eq:dsigdmu}]. For low $E_\nu$ and correspondingly low energy transfer, the factor $(E_\nu-\omega)^2$ would enhance the scattering cross section slightly (lower and upper panels of Fig.~\ref{fig:Sig}), mainly due to negative values of $\omega$ (see the middle panels of Fig.~\ref{fig:Sig}).  
Since $S_{V,A}(q, -\omega) = e^{-\omega/T}S_{V,A}(q, \omega)$, the structure factors with positive $\omega$ becomes more dominant with higher values of $\omega$. Therefore, for high $E_\nu$ with high energy transfer, the factor $(E_\nu-\omega)^2$ suppresses the cross section (lower panels of Fig.~\ref{fig:Sig}).
The second limitation is about the range of $(q, \omega)$ to be integrated over. For neutrino-nucleon scattering, the kinetically allowed 
range is $-q\le \omega \le {\rm min}[2E_\nu-q, q]$ for a given $q$, 
while in the definition of $S_{V,A}(q)$, all the range of $-\infty < \omega < \infty$ has been taken into account [Eq.~\eqref{eq:sva_def}]. This effect further reduces the cross sections at high $E_\nu$ with the use of dynamic structure factors. Compared to the calculations based on the static structures, the use of the dynamic structure factors can enhance (suppress) the total cross sections for low (high) $E_\nu$ by 10\%--20\%.  

We also compare our results with those using $S_{V,A}(q, \omega)$ of Bedaque {\it et al} \cite{Bedaque_2018}.
As illustrated in the upper panels of Fig.~\ref{fig:Sig}, our $d\sigma/d\cos\theta$ are slightly larger and evidently larger at conditions with $z\approx 0.23$ and $z\approx 0.62$, respectively.  
The primary reason for this discrepancy is that our $S_A(q, \omega)$ is normalized to a larger $S_A(q)$ [and $S_A(0)$]. Note that for $z\approx 0.62$, the BHF calculation of $S_A(0)$ should be more preferred than those calculated via the virial expansion as adopted in \cite{Bedaque_2018}. If we instead normalize $S_A(q, \omega)$ to that of \cite{Bedaque_2018}, almost the same $d\sigma/d\cos\theta$ and total cross section are obtained (see the magenta and green lines in the upper panels of Fig.~\ref{fig:Sig}). This suggests that the angular cross section and the total cross section are not very sensitive to the detailed distribution of $S_{V,A}(q, \omega)$ in $\omega$, and the method we use to construct the dynamic structure factors using sum rules is effective.   
The middle panels of Fig.~\ref{fig:Sig} compare the energy differential cross sections $d\sigma/d\omega$.  Comparing to the non-interacting case, we find that neutron-neutron correlation mainly reduces $d\sigma/d\omega$ at small $\omega$. The discrepancies between our results with those of \cite{Bedaque_2018} arise from differences in $S_{V,A}(q, \omega)$. Note that at $n=0.0126$~fm$^{-3}$, the overly suppressed $S_{A}(q, \omega)$ from \cite{Bedaque_2018} results in a dip in $d\sigma/d\omega$ around $\omega=0$ (the middle right panel of Fig.~\ref{fig:Sig}).
We find that when employing our $S_{A}(q, \omega)$ normalized to $S_A(q)$ of \cite{Bedaque_2018}, the $d\sigma/d\omega$, although still exhibiting notable differences, get closer to those obtained from $S_A(q, \omega)$ of \cite{Bedaque_2018}.    

In the RPA results presented in Fig.~\ref{fig:Sig}, we use the RPA $S_A(q, \omega)$ from \cite{Burrows:1998cg} and the $S_V(q, \omega)$ for a non-interacting neutron gas to compute the differential and the total cross sections [Eq.~\eqref{eq:dsigdwdmu}]. At $n=0.004$~fm$^{-3}$, the RPA yields different $d\sigma/d\cos\theta$ but similar $d\sigma/d\omega$ when compared to our results. As the RPA spin structure factor is higher and meanwhile, the density structure factor for a free neutron gas is lower that those of ours, the resulting total cross sections derived are similar to ours (the lower left panel of Fig.~\ref{fig:Sig}). At a higher density, the density structure factor becomes more important, and our total cross sections using enhanced $S_V(q, \omega)$ are higher than the RPA calculations.

\section{Discussion and Summary} 
\label{sec:sum}

In this work, we have calculated the static structure factors in the long wavelength limit for pure neutron matter at subsaturation densities using the BHF approach.
The obtained results are generally consistent with those from the virial expansion at low and intermediate densities. Moreover, the BHF results are believed to be reliable at high densities, extending beyond the applicability of the virial expansion. We find that $S_A(0)$ decreases monotonically with density, while $S_V(0)$ increases firstly and then decreases. Consequently, the neutrino-neutron scattering cross section is suppressed.
The numerical results at selected values of temperatures and densities are also accessible as Supplemental Material \cite{supp.mat.}. Moreover, we have also provided analytical fits to $z$ and $S_{V,A}(0)$.

Due to a large scattering length and a relatively short effective range, the dilute neutron gas could resemble a unitary gas and thereby shares similar properties in terms of the static and the dynamic structure factors. By extrapolating the behaviors of $S_{V,A}(q)$ for a unitary gas at supernova conditions, we determine $S_{V,A}(q)$ for the neutron gas using $S_{V,A}(0)$ obtained from either the BHF approach or the virial expansion. Additionally, we introduce a phenomenological approach to derive the dynamic structure factors with simple analytical expressions based on the sum rules.           

The in-medium neutrino-neutron scattering cross sections are calculated using the obtained $S_{V,A}(0)$, $S_{V,A}(q)$, and $S_{V,A}(q, \omega)$. Overall, all the cross sections are suppressed compared to the one for a non-interacting neutron gas by 10\%--20\%.
As $S_{V,A}(q)$ vary slowly at small $q$, relevant to neutrino-nucleon scattering, the cross sections using $S_{V,A}(0)$ and $S_{V,A}(q)$ are very similar to each other. When the dynamic structure factors are used, the differential and the total cross section could be altered by 10\%--20\%, depending on the energy of the incoming neutrino. 
These modifications, though relatively small, could significantly impact the dynamics of CCSNe.   

The study presented in this work can be improved in many aspects. For a more comprehensive study, the three-body forces need to be considered for our BHF calculation, even though
their impact is expected to be modest at subsaturation densities near the neutrinosphere.
Our reliance on the similarities between the neutron gas and a unitary gas to derive the structure factors should be validated through more independent studies such as \cite{Bedaque_2018}.
We have assumed that the dynamic structure factors can be described by a double-Gaussian-peak structure and that $S_A$ obeys the $f$-sum rule as $S_V$. These assumptions will fail at high-density regions since the nucleon-nucleon interaction can violate the normalization condition for $S_A(q,\omega)$ given by Eq.~(\ref{eq:fsum}) and enhance $S_A(q, \omega)$ at high $\omega$ due to multiparticle excitations. Therefore, the dynamic spin structure factor at high densities and high $\omega$ may require a separate treatment \cite{Guo:2019cvs,Raffelt:2001kv}. The exploration of other many-body methods to derive the static or dynamic structure factors is also highly encouraged. For instance, a recent work considering both mean field effects and RPA correction using the $\chi$EFT potential is a step in this direction \cite{Shin:2023sei}. The incorporation of full dynamic structure factors for neutrino transport in CCSN simulations is highly desired in future studies. 
The study in this work is limited to pure neutron matter. To apply the results to CCSN  simulations with a wide range of conditions, it will be important to include neutrino-proton interactions and extend similar investigations to asymmetric nuclear matter \cite{Horowitz:2016gul}. We plan to work on these aspects in the future.

\begin{acknowledgments}
GG acknowledges support by the National Natural Science Foundation of China (No.~12205258), the Fundamental Research Funds for the
Central Universities, China University of Geosciences (Wuhan) (No.~CUG240630), and the ``CUG Scholar" Scientific
Research Funds at China University of Geosciences (Wuhan) (No.~2021108).
MRW acknowledges support by the National Science and Technology Council (No.~111-2628-M-001-003-MY4), the Academia Sinica (No.~AS-CDA-109-M11), and the Physics Division of the National Center for Theoretical Sciences, Taiwan. GMP acknowledges support by the European Research Council (ERC) under the European Union's Horizon 2020 research and innovation programme (ERC Advanced Grant KILONOVA No.~885281),  the Deutsche Forschungsgemeinschaft (DFG, German Research Foundation) - Project-ID 279384907 - SFB 1245, and MA 4248/3-1, and the State of Hesse within the Cluster Project ELEMENTS.
\end{acknowledgments}

\end{document}